\documentclass[conference]{IEEEtran}
\IEEEoverridecommandlockouts

\usepackage{cite}
\usepackage{amsmath,amssymb,amsfonts}
\usepackage{algorithmic}
\usepackage{graphicx}
\usepackage{textcomp}
\usepackage{xcolor}
\def\BibTeX{{\rm B\kern-.05em{\sc i\kern-.025em b}\kern-.08em
    T\kern-.1667em\lower.7ex\hbox{E}\kern-.125emX}}
\begin{document}

\title{
{Beyond Spherical Wavefront: Near-Field Channel Estimation Under Wavefront Anisotropy}
\thanks{The research presented in this paper has been kindly supported by the projects as follows, National Natural Science Foundation of China under Grants (62394294, 62394290), the Fundamental Research Funds for the Central Universities under Grant 2242022k60006. }
}

\author{
    \IEEEauthorblockN{Heling Zhang\IEEEauthorrefmark{1}\IEEEauthorrefmark{3} Xiujun Zhang\IEEEauthorrefmark{2} Xiaofeng Zhong\IEEEauthorrefmark{1}\IEEEauthorrefmark{3} Shidong Zhou\IEEEauthorrefmark{1}\IEEEauthorrefmark{3}}
    \IEEEauthorblockA{\IEEEauthorrefmark{1}Department of Electronic Engineering, Tsinghua University, Beijing, China}
    \IEEEauthorblockA{\IEEEauthorrefmark{2}Beijing National Research Center for Information Science and Technology}
    \IEEEauthorblockA{\IEEEauthorrefmark{3}State Key Laboratory of Space Network and Communications\\
    Emails: zhanghl24@mails.tsinghua.edu.cn, \{zhangxiujun, zhongxf, zhousd\}@tsinghua.edu.cn}
}

\maketitle

\begin{abstract}
Extremely large aperture arrays (ELAAs) and millimeter-wave (mmWave) technologies are essential for achieving high data rates in future wireless communication systems. To perform precise beamforming, these systems require accurate channel estimation, in which the near-field wavefront curvature effect must be taken into account. Existing channel estimation methods rely on the spherical wavefront channel (SWC) model, which is suitable for near-field propagation with point sources, scatterers, and reflection planes. However, when a near-field curved reflecting surface exists, the wavefront of the reflected wave becomes anisotropic rather than spherical, causing the SWC model to become inaccurate. To address this problem, in this paper, we formulate a parameterized model for the anisotropic wavefront channel (AWC). Using this model, we propose a channel estimation algorithm based on physical parameter recovery for the AWC. Simulation results reveal that the AWC no longer retains sparsity in the angle-distance domain. Furthermore, the results demonstrate how different physical characteristics of the propagation scenario affect the degree of wavefront anisotropy, and confirm the effectiveness of our proposed algorithm in AWC scenarios.
\end{abstract}


\section{Introduction}
The impending transition to 6G wireless systems demands unprecedented data rates, minimal latency, and strict reliability. To meet these ambitious goals, extremely large aperture arrays (ELAAs) and millimeter-wave (mmWave) have been recognized as foundational technologies. Advancing the legacy of 5G massive MIMO, ELAA drastically multiplies the number of antenna elements to deliver highly precise beamforming and substantial spectral efficiency improvements. Meanwhile, the exploitation of the mmWave band provides the abundant available bandwidth for massive capacity expansion. Furthermore, the millimeter-level wavelength permits highly compact antenna spacing, which facilitates the practical deployment of ELAAs within restricted physical spaces \cite{ref1}.

Accurate channel estimation is essential for precise precoding and beamforming. For such ELAA-mmWave systems, wavefront curvature must be appropriately modeled and estimated. Existing methods rely on the sparse spherical wavefront channel (SWC) model, where the signal reaches the base station (BS) array through only a few paths, each exhibiting a spherical wavefront. Therefore, each path is fully defined by its path gain, direction of arrival, and propagation distance, which determines the wavefront curvature \cite{ref2,ref3}. Due to the small number of paths, the channel is highly sparse in the angle-distance (polar) domain, allowing algorithms to estimate the channel via sparse recovery. For example, \cite{ref4} and \cite{ref5} discretely sample the polar domain to build a steering vector dictionary and recover the channel using orthogonal matching pursuit (OMP). To reduce the dimension of the sampled grids, some algorithms perform matching solely in the angular domain. For instance, \cite{ref6, ref7, ref17} treat the distance as an iterative parameter, updating the distance estimate alongside the angular matching in each step. Meanwhile, \cite{ref8} extracts the propagation delay of each path from wideband observation in the frequency domain, thereby avoiding the distance search.

However, when curved reflecting surfaces exist in the propagation environment, the reflected waves exhibit wavefront anisotropy, meaning their curvatures differ in different directions \cite{ref9}. This anisotropy alters the phase distribution of the channel response across the array. Consequently, the anisotropic wavefront channel (AWC) deviates from the SWC model, losing its sparsity in the polar domain and thereby causing the performance of existing estimation methods to degrade. To address this issue, our main contributions are summarized as follows:
\begin{itemize}
    \item We formulate a parameterized model for the single-path AWC. Through simulations, we reveal the non-sparsity of AWC in the polar domain, and how specific physical characteristics of the propagation environment determine the degree of wavefront anisotropy.
    \item We propose a channel estimation algorithm for the AWC based on the aforementioned channel model.
\end{itemize}
\textit{Notation}: $(\cdot)^{-1}$, $(\cdot)^{T}$, and $(\cdot)^{H}$ denote the inverse, transpose, and conjugate transpose. $[\mathbf{A}]_{i,j}$ denotes the entry in the $i$-th row and $j$-th column of matrix $\mathbf{A}$, while $[\mathbf{A}]_{i_1:i_2, j_1:j_2}$ denotes the submatrix spanning rows $i_1$ to $i_2$ and columns $j_1$ to $j_2$. $\text{unvec}(\cdot)$ 
reshapes vector into a matrix.

\section{System Model}\label{sec: System Model}
\begin{figure}[!t]
\centering
\includegraphics[width=\linewidth]{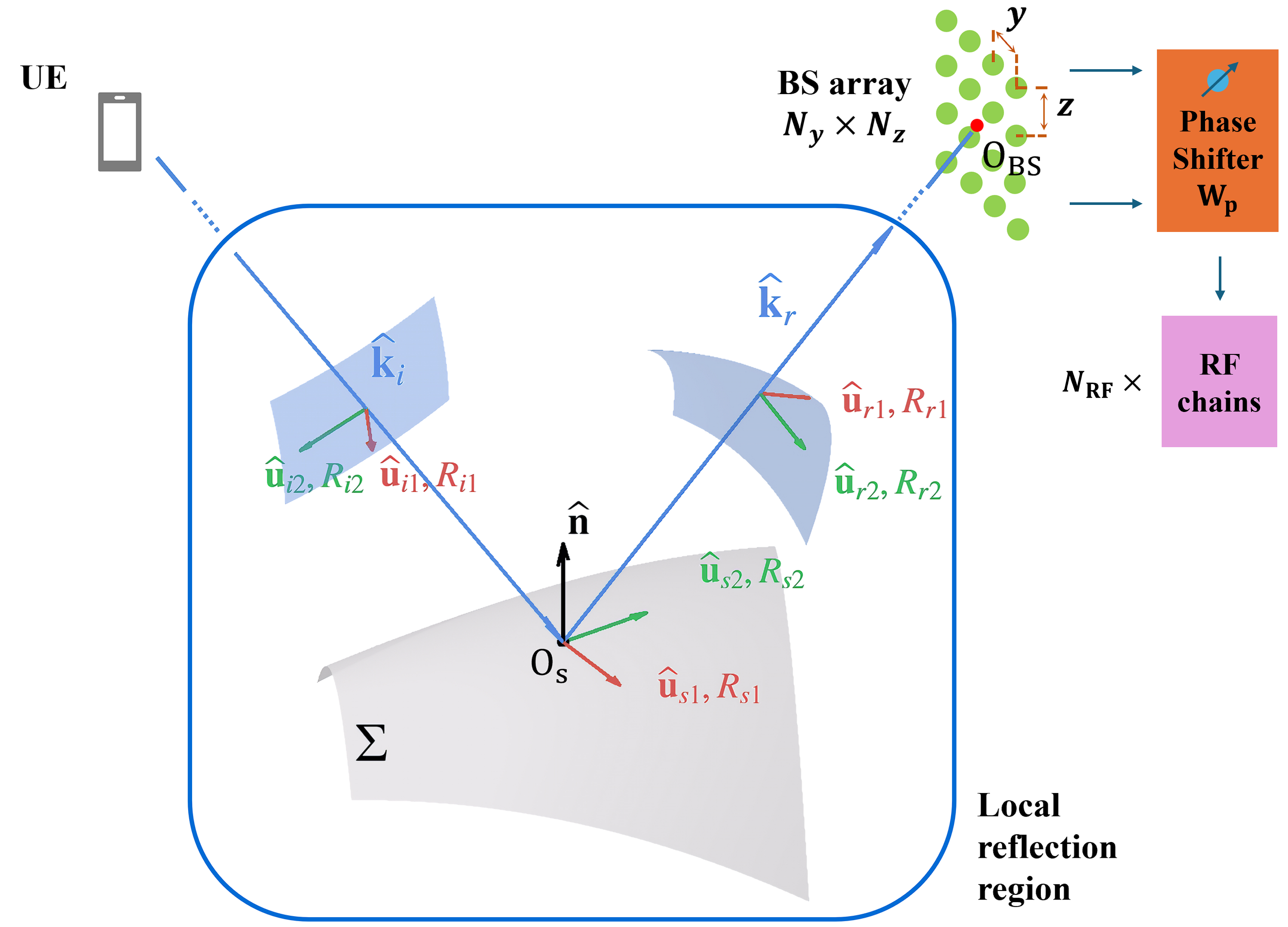}
\caption{Considered propagation scenario.}
\label{img:curved_reflection}
\end{figure}

As in Fig. \ref{img:curved_reflection}, we consider a narrowband ELAA system with hybrid beamforming. The BS is equipped with a uniform planer array (UPA) having $N_y\times N_z= N$ antennas. These antennas are connected to $N_{\text{RF}}$ RF chains via an analog phase-shifting network. The carrier frequency is $f$, the wavelength is $\lambda$, and the antenna spacing in both dimensions is $\lambda/2$. The user equipment (UE) is equipped with a single antenna. The environment contains a single scatterer with a smooth curved reflecting surface, which provides the sole reflection path for the channel. The principal radii of scatterer are much larger than the carrier wavelength.

In the uplink channel estimation scheme, the UE consecutively transmits $P$ unit pilot symbols to the BS. The signal received by BS during the $p$-th pilot transmission is given by
\begin{equation}
\mathbf{y}_p = \mathbf{W}_p (\mathbf{h} + \mathbf{n}_p),
\end{equation}
where $\mathbf{h} = a\mathbf{c}(\bar{\mathbf{k}},\bar{\mathbf{Q}})$ denotes the AWC channel response vector. The AWC steering vector $\mathbf{c}(\bar{\mathbf{k}},\bar{\mathbf{Q}})$ is of constant envolope 1, and the phase shifts are characterized with parameters $\bar{\mathbf{k}}$ and $\bar{\mathbf{Q}}$. Distinct from the SWC, where the wavefront is isotropic and the curvature is depicted by a single parameter, in AWC, a matrix $\bar{\mathbf{Q}}$ captures the effective anisotropic curvature across the BS array aperture. $\bar{\mathbf{k}}$ corresponds to the wave vector associated with the angle of arrival of the incident AWC path and $a$ stands for the path gain. The formulation of the AWC steering vector is detailed in Section \ref{sec: AWC Formulation}.

Meanwhile, $\mathbf{W}_p \in \mathbb{C}^{N_\text{RF}\times N}$ is the hybrid beamforming matrix at the $p$-th pilot, and $\mathbf{n}_p \sim \mathcal{CN}(\mathbf{0}, \mathbf{I}) $ is the additive receiver noise. By stacking the observations from all $P$ pilots, we obtain the total observation model:
\begin{equation}\label{eq:observation model}
\mathbf{y} = \mathbf{W} \mathbf{h} + \mathbf{n},
\end{equation}
where $\mathbf{y} = [\mathbf{y}_1^T,\dots,\mathbf{y}_p^T]^T$, $\mathbf{W} = [\mathbf{W}_{1}^T,\dots,\mathbf{W}_{p}^T]^T$ and $\mathbf{n} = [(\mathbf{W}_{1}\mathbf{n}_{1})^T,\dots,(\mathbf{W}_{p}\mathbf{n}_{p}^T)]^T$.

Specifically, since the combining matrix $\mathbf{W}$ can be flexibly designed, we configure it as a subsampled randomized Fourier transform (SRFT) matrix. This choice is motivated by the spectrum-preserving property of the projection operator $\mathbf{W}^H\mathbf{W}$ \cite{ref10}, meaning that the spectrum of $\mathbf{W}^H\mathbf{y}$ retains the similar features of the spectrum of $\mathbf{h}$. Furthermore, $\mathbf{W}^H\mathbf{y}$ can be efficiently computed via the fast Fourier transform (FFT). Additionally, under this design, the noise vector $\mathbf{n}$ in (\ref{eq:observation model}) remains white, thereby eliminating the need for pre-whitening.

\section{AWC Formulation}\label{sec: AWC Formulation}
In this section, we formulate the AWC steering vector $\mathbf{c}(\bar{\mathbf{k}},\bar{\mathbf{Q}})$ introduced in Section \ref{sec: System Model}. While the geometric formulation of the anisotropic wavefront strictly follows Section III of \cite{ref9}, we specifically extend its framework to derive the spatial response for ELAA channels.

\textit{Preliminary}: The anisotropic curvature of a wavefront or a reflecting surface is described by a curvature matrix $\mathbf{Q}$. Defining the propagation direction or the normal as the $h$-axis with the origin at the wavefront vertex, the surface can be approximated within a neighborhood of the origin by the quadratic form $h = -\frac{1}{2}\mathbf{x}^T\mathbf{Q}\mathbf{x}$, where $\mathbf{x} \in \mathbb{R}^2$ is the projection vector onto a reference plane orthogonal to the $h$-axis. The curvature matrix $\mathbf{Q}$ is determined by the chosen coordinate bases on this plane. If a basis is selected such that $\mathbf{Q}$ is diagonalized, these basis directions are termed the principal directions of the surface, with the corresponding diagonal entries representing the principal curvatures. The principal radii of curvature $R_1, R_2$ are given by the reciprocals of these principal curvatures.

Firstly, to characterize curvature matrix $\mathbf{Q}_\text{BS}$ at the BS array center, we evaluate the wavefront curvatures at the point of reflection, which is described by the curvature matrix $\mathbf{Q}_\text{r}$ and its corresponding coordinate bases $(\hat{\mathbf{u}}_{\text{r},1}, \hat{\mathbf{u}}_{\text{r},2})$.

$\mathbf{Q}_\text{r}$ are governed by \cite{ref9}:
\begin{equation} \label{eq:Deschamps}
    \mathbf{Q}_\text{r} = \mathbf{Q}_\text{i} + 2(\mathbf{\Theta}^{-1})^T \mathbf{Q}_\text{s} \mathbf{\Theta}^{-1} \cos \theta_\text{i},
\end{equation}
where $\mathbf{Q}_\text{i}$ and $\mathbf{Q}_\text{s}$ are the curvature matrices of the incident wave and the reflecting surface, defined in their respective local coordinate bases $(\hat{\mathbf{u}}_{\text{i},1}, \hat{\mathbf{u}}_{\text{i},2})$ and $(\hat{\mathbf{u}}_{\text{s},1}, \hat{\mathbf{u}}_{\text{s},2})$. $\mathbf{\Theta}$ is a projection matrix that aligns $\mathbf{Q}_\text{i}$ with the surface coordinate system, defined as
\begin{equation}
\mathbf{\Theta} = \begin{bmatrix} \hat{\mathbf{u}}_{\text{i},1}^T \hat{\mathbf{u}}_{\text{s},1} & \hat{\mathbf{u}}_{\text{i},1}^T \hat{\mathbf{u}}_{\text{s},2} \\ \hat{\mathbf{u}}_{\text{i},2}^T \hat{\mathbf{u}}_{\text{s},1} & \hat{\mathbf{u}}_{\text{i},2}^T \hat{\mathbf{u}}_{\text{s},2} \end{bmatrix}.
\end{equation}
$\theta_i$ denotes the incidence angle relative to the scatterer surface.

Subsequently, $(\hat{\mathbf{u}}_{\text{r},1}, \hat{\mathbf{u}}_{\text{r},2})$ are determined according to the law of reflection \cite{ref9}:
\begin{equation}
\hat{\mathbf{u}}_{\text{r},1} = \hat{\mathbf{u}}_{\text{i},m} - 2(\hat{\mathbf{n}}_\text{norm}^T \hat{\mathbf{u}}_{\text{i},m})\hat{\mathbf{n}}_\text{norm}, \quad m \in \{1, 2\},
\end{equation}
where $\hat{\mathbf{n}}_\text{norm}$ is the normal of the reflecting surface.

As the wave propagates a distance $s_{\text{r}}$ from the reflection point to the BS array, the curvature matrix $\mathbf{Q}_r$ also evolves to $\mathbf{Q}_\text{BS}$. Both principal radii of curvature of the wavefront increase by $s_{\text{r}}$, while their principal directions remain unchanged. This relationship is described as below \cite{ref9}:
\begin{equation} \label{eq:Deschamps2}
    \mathbf{Q}_{\text{BS}} = (\mathbf{Q}_{\text{r}}^{-1} + s_\text{r} \mathbf{I})^{-1}.
\end{equation}

By applying the distance formulas from \cite{ref9} based on $\mathbf{Q}_\text{BS}$ and its coordinate basis, we calculate the propagation distance to each array element, thereby extending the framework of \cite{ref9} to construct the phase response of the whole BS array. Let $S(\mathbf{p}_{n_y, n_z})$ denote the distance to the $(n_y, n_z)$-th element, where $\mathbf{p}_{n_y, n_z}$ is the position vector from the array center to this element. The distance function $S(\mathbf{p})$ can be calculated using a Taylor expansion around the center of BS array \cite{ref9}:
\begin{equation}
S(\mathbf{p}) \approx S(\mathbf{0}) + \hat{\mathbf{k}}_r^T \mathbf{p} + \frac{1}{2} \mathbf{p}^T \mathbf{H} \mathbf{p}.
\end{equation}
The Hessian matrix is given by $\mathbf{H} = \mathbf{U}_\text{r}^T \mathbf{Q}_\text{BS} \mathbf{U}_\text{r}$, where $\mathbf{U}_\text{r}=[\hat{\mathbf{u}}_{\text{r},1}, \hat{\mathbf{u}}_{\text{r},2}]$ represents the projection matrix onto the array plane, and the gradient $\hat{\mathbf{k}}_r$ is given by the unit direction vector of the reflected wave.

The phase of response at the $(n_y, n_z)$-th antenna element is then determined by the propagation distance:
\begin{equation}
[\tilde{\mathbf{C}}]_{n_y,n_z} = \tilde{a} e^{-j\frac{2\pi}{\lambda}S(\mathbf{p}_{n_y, n_z})},
\end{equation}
where $\tilde{\mathbf{C}}$ is the unvectorized form of the unnormalized AWC steering vector, and the path gain $\tilde{a}=\frac{e^{j2\pi S(\mathbf{0})/\lambda}}{\sqrt{N}}a$ is determined by factors such as propagation distance, path loss, polarization, and the reflector's material properties. For the UPA, the position vector of an element is given by $\mathbf{p}_{n_y, n_z} = \delta_{n_y} d_\text{ant} \hat{\mathbf{y}} + \delta_{n_z} d_\text{ant} \hat{\mathbf{z}}$, where $d_\text{ant}$ denotes the antenna spacing along both array axes, and $\delta_{n_y} = n_y - (N_y+1)/2$, $\delta_{n_z} = n_z - (N_z+1)/2$ are the zero-centered antenna indices. Substituting $\mathbf{p}_{n_y, n_z}$ into the Taylor expansion of $S(\mathbf{p})$, and subsequently into the response equation, yields the following simplified expression:
\begin{equation}\label{eq: effective AWC steering matrix}
[\mathbf{C}]_{n_y, n_z} = \frac{1}{\sqrt{N}}e^{-j 2\pi \left( \bar{\mathbf{k}}^T \mathbf{n}_{y,z} + \frac{1}{2} \mathbf{n}_{y,z}^T \bar{\mathbf{Q}} \mathbf{n}_{y,z} \right)},
\end{equation}
where the index vector $\mathbf{n}_{y,z}=[\delta_{n_y},\delta_{n_z}]^T$. $\mathbf{c}(\bar{\mathbf{k}},\bar{\mathbf{Q}}) = \text{unvec}(\mathbf{C})$ denotes the normalized AWC steering vector. In this expression, the effective direction $\bar{\mathbf{k}}$ and the curvature matrix $\bar{\mathbf{Q}}$ are jointly determined by the relative wave propagation direction, the array orientation, and the principal curvatures and directions of the wavefront.

Specifically, by setting the reflector's curvature matrix to $\mathbf{Q}_s = 0\mathbf{I}$, or by taking the limit $\mathbf{Q}_s = q\mathbf{I}$ as $q \to \infty$, the proposed model reduces to two distinct SWC steering vectors. These correspond to a planar reflector and an isotropic point scatterer, respectively. This demonstrates that the traditional SWC model is a special case of the AWC model.

\section{Proposed Channel Estimation Algorithm}\label{sec: Proposed Channel Estimation Algorithm}
Based on the AWC model established in Section \ref{sec: AWC Formulation}, channel estimation for $\mathbf{h}=a\mathbf{c}(\bar{\mathbf{k}},\bar{\mathbf{Q}})$ can be formulated as a parameter estimation problem for the path geometric parameters $\bar{\mathbf{k}}$, $\bar{\mathbf{Q}}$, and the path gain $a$. In the proposed channel estimation algorithm, we first recover a coarse channel estimate $\hat{\mathbf{h}}$ from the observations. From this recovered signal, we extract the curvature parameter $\bar{\mathbf{Q}}$ based on phase differencing, followed by the direction parameter $\bar{\mathbf{k}}$. Following a joint refinement of these parameters, the optimized results are used to evaluate the path gain $a$ and yield the final channel estimate. Hereafter, we use $\mathbf{H} = \text{unvec}(\mathbf{h})$ and $\hat{\mathbf{H}} = \text{unvec}(\hat{\mathbf{h}})$ to denote the channel response, and its coarse estimate, arranged according to the antenna array geometry. 

\subsection{Recovery of Channel Steering Vector}
According to (\ref{eq: effective AWC steering matrix}), if a 2D FFT is applied to $\mathbf{H}$, its spectral energy will be highly concentrated within a subspace $\mathcal{U}$ spanned by the columns of an orthogonal Fourier basis matrix $\mathbf{U}$. Therefore, the observation model in (\ref{eq:observation model}) is rewritten as:
\begin{equation}
\mathbf{y} \approx \mathbf{W}\mathbf{U}\tilde{\mathbf{h}} + \mathbf{n}.
\end{equation}
Because the energy of $\mathbf{h}$ is highly concentrated, the dimension of $\tilde{\mathbf{h}}$ is smaller than that of $\mathbf{y}$, allowing $\tilde{\mathbf{h}}$ to be recovered via least squares (LS). Owing to the spectrum-preserving property of $\mathbf{W}$, the term $\mathbf{U}^H(\mathbf{W}^H\mathbf{W}\mathbf{U})$ approximates an identity matrix. Thus, the LS recovery of $\mathbf{h}$ can be expressed as:
\begin{equation}
\hat{\mathbf{h}} = \mathbf{U}\tilde{\mathbf{h}} \approx \mathbf{U}\mathbf{U}^H\mathbf{W}^H\mathbf{y},
\end{equation}
where $\mathbf{U}\mathbf{U}^H$ serves as the projection operator onto $\mathcal{U}$.

To determine this subspace, we leverage the spectrum-preserving property of $\mathbf{W}^H\mathbf{W}$. Since the energy of $\mathbf{h}$ is concentrated within $\mathcal{U}$, the energy of $\mathbf{W}^H\mathbf{y}$ must also be concentrated in this identical subspace. Consequently, we first apply a 2D FFT to the unvectorized $\mathbf{W}^H\mathbf{y}$ to locate its energy-concentrated region, which effectively identifies the bases for $\mathbf{U}$. Subsequently, we set the frequency components of $\mathbf{W}^H\mathbf{y}$ outside this region to zero. This operation is equivalent to projecting $\mathbf{W}^H\mathbf{y}$ onto $\mathcal{U}$. Through this process, we obtain the recovered channel response $\hat{\mathbf{h}}$.

The energy-concentrated region of $\mathbf{W}^H\mathbf{y}$ is located as below. Firstly, the 2D power spectrum of $\mathbf{W}^H\mathbf{y}$ is smoothed and we subtract its mean plus three standard deviations from the smoothed spectrum. Subsequently, Kadane's algorithm \cite{ref11} is executed along both dimensions to extract the maximum-energy rectangular bounding box.

\subsection{Phase-difference-based Parameter Extraction}

As indicated in (\ref{eq: effective AWC steering matrix}), the curvature parameters are embedded within the quadratic phase components. Consequently, we employ a phase-difference mechanism to transform this quadratic phase into a 2D single-frequency structure, thereby enabling the extraction of the curvature parameters by estimating their corresponding frequencies. Specifically, we extract a reference sub-matrix $\mathbf{\hat{H}}^{(1)}$ from the top-left of the reconstructed response $\mathbf{\hat{H}}$ and its shifted counterpart $\mathbf{\hat{H}}^{(2)}$ from the bottom-right:
\begin{equation}
\begin{aligned}
\mathbf{\hat{H}}^{(1)} &= [\mathbf{\hat{H}}]_{1:N_y-\Delta_y, 1:N_z-\Delta_z}\quad\\ \mathbf{\hat{H}}^{(2)} &= [\mathbf{\hat{H}}]_{1+\Delta_y:N_y, 1+\Delta_z:N_z}.
\end{aligned}
\end{equation}
Their element-wise conjugate multiplication yields the first differential matrix:
\begin{equation}
\begin{aligned}
    \relax[\boldsymbol{\Psi}_1]_{n_y, n_z} &= \left([\hat{\mathbf{H}}^{(1)}]_{n_y, n_z}\right)* [\hat{\mathbf{H}}^{(2)}]_{n_y, n_z} \\
    &= |\tilde{a}|^2 e^{j 2\pi (f_{1,y} \delta_{n_y} + f_{1,z} \delta_{n_z} + \Phi_1)} + [\mathbf{N}_1]_{n_y, n_z},
\end{aligned}
\end{equation}
where $\mathbf{N}$ denotes the equivalent noise term resulting from the original receiver noise, the recovery of channel response, and the difference operation. $\Phi_1$ is a constant phase.

Analogously, by differencing the sub-matrices extracted from the top-right and bottom-left corners of $\mathbf{\hat{H}}$, we obtain the second differential matrix:
\begin{equation}
    [\boldsymbol{\Psi}_2]_{n_y, n_z} = |\tilde{a}|^2 e^{j 2\pi (f_{2,y} \delta_{n_y} + f_{2,z} \delta_{n_z} + \Phi_2)} + [\mathbf{N}_2]_{n_y, n_z}.
\end{equation}
The frequencies $\{f_{1,y}, f_{1,z}, f_{2,y}, f_{2,z}\}$ embedded within $\boldsymbol{\Psi}_1$ and $\boldsymbol{\Psi}_2$ are uniquely determined by the curvature matrix $\bar{\mathbf{Q}}_k$, thereby allowing for the reconstruction of its parameters. 

To extract the target frequencies, we compute the 2D FFT of $\boldsymbol{\Psi}_1$ and $\boldsymbol{\Psi}_2$ and perform a joint peak search over their power spectra. Crucially, these frequencies are not independent but adhere to the following linear constraint:
\begin{equation}
\Delta_y f_{1,y} - \Delta_z f_{1,z} = \Delta_y f_{2,y} + \Delta_z f_{2,z} = V
\end{equation}
This implies that for any specified intercept $V$, the potential frequency candidates are restricted to a 1D line trajectory on each 2D spectrum. The true intercept $V^*$ corresponds to the specific trajectories that simultaneously intersect the genuine spectral peaks in both spectra. Let $P_1(f_y, f_z)$ and $P_2(f_y, f_z)$ denote the power spectra of $\boldsymbol{\Psi}_1$ and $\boldsymbol{\Psi}_2$, respectively, and $\mathcal{L}_i(V)$ denote the frequency line dictated by $V$ on spectrum $P_i$. We traverse the feasible range of $V$ to identify $V^*$ by solving the following max-min optimization:
\begin{equation}
V^* = \mathop{\arg\max}\limits_{V} \min_{i \in {1,2}} \left\{ \max_{(f_y, f_z) \in \mathcal{L}_i(V)} P_i(f_y, f_z) \right\}
\end{equation}
Once $V^*$ is determined, the target frequencies are pinpointed as the coordinates exhibiting the maximum energy along the corresponding optimal lines. Finally, with these extracted frequencies, $\hat{\bar{\mathbf{Q}}}$ can be explicitly solved via the phase-difference-based relations as below:
\begin{equation}\label{eq:quadratic params}
\begin{aligned}
    \relax[\hat{\bar{\mathbf{Q}}}]_{1,1} &= \frac{-f_{1,y} - f_{2,y}}{2\Delta_y} \\
    [\hat{\bar{\mathbf{Q}}}]_{2,2} &= \frac{-f_{1,z} + f_{2,z}}{2\Delta_z} \\
    [\hat{\bar{\mathbf{Q}}}]_{1,2} &= [\hat{\bar{\mathbf{Q}}}]_{2,1} = \frac{1}{2} \left( \frac{-f_{1,y} + f_{2,y}}{2\Delta_z} + \frac{-f_{1,z} - f_{2,z}}{2\Delta_y} \right)
\end{aligned}
\end{equation}

By exploiting the estimated $\hat{\bar{\mathbf{Q}}}$, the energy spread caused by the quadratic phase can be eliminated to estimate $\bar{\mathbf{k}}$. We construct a phase compensation matrix $\mathbf{S}$, whose elements are $[\mathbf{S}]_{n_y, n_z} = e^{-j\pi\left(\mathbf{n}_{y,z}^T \hat{\bar{\mathbf{Q}}} \mathbf{n}_{y,z} \right)}$, and compute $\mathbf{S}^* \odot \mathbf{\hat{H}}$. This multiplication yields a residual signal that is merely a superposition of a single-frequency component and noise. Therefore, the direction parameter $\hat{\bar{\mathbf{k}}}$ can be directly acquired by computing its 2D power spectrum and identifying the global peak.

\subsection{Parameter Refinement and Channel Reconstruction}
To further refine the parameters, we employ the Levenberg-Marquardt (LM) algorithm \cite{ref12} utilizing the acquired estimates as initializations. Specifically, we minimize the mean squared error (MSE) between the true and the reconstructed observations:
\begin{equation}
    \min_{\bar{\mathbf{k}}, \bar{\mathbf{Q}}, a} \left\| \mathbf{y} - a\mathbf{W}\mathbf{c}(\bar{\mathbf{k}}, \bar{\mathbf{Q}}) \right\|_2^2
\end{equation}
In each iteration, the path gain $a$ is first computed via LS. Subsequently, holding $a$ fixed, the Jacobian matrix of the residual with respect to $\bar{\mathbf{k}}$ and $\bar{\mathbf{Q}}$ is evaluated to update these parameters. Upon convergence, the optimized parameters $a^*$, $\bar{\mathbf{k}}^*$, and $\bar{\mathbf{Q}}^*$ are utilized to reconstruct the final channel estimate as $\hat{\mathbf{h}}^* = a^*\mathbf{c}(\bar{\mathbf{k}}^*, \bar{\mathbf{Q}}^*)$.
\section{Simulation Results}\label{sec: Simulation Results}
In this section, we first demonstrate the non-sparsity of the AWC in the polar domain caused by wavefront anisotropy, alongside the impact of propagation scenario characteristics on the degree of this anisotropy. Subsequently, we verify the effectiveness of the proposed algorithm in AWCs.

\subsection{Non-Sparsity of AWC in the Polar Domain}

\begin{figure}[htbp]
    \centering
    \includegraphics[width=\linewidth]{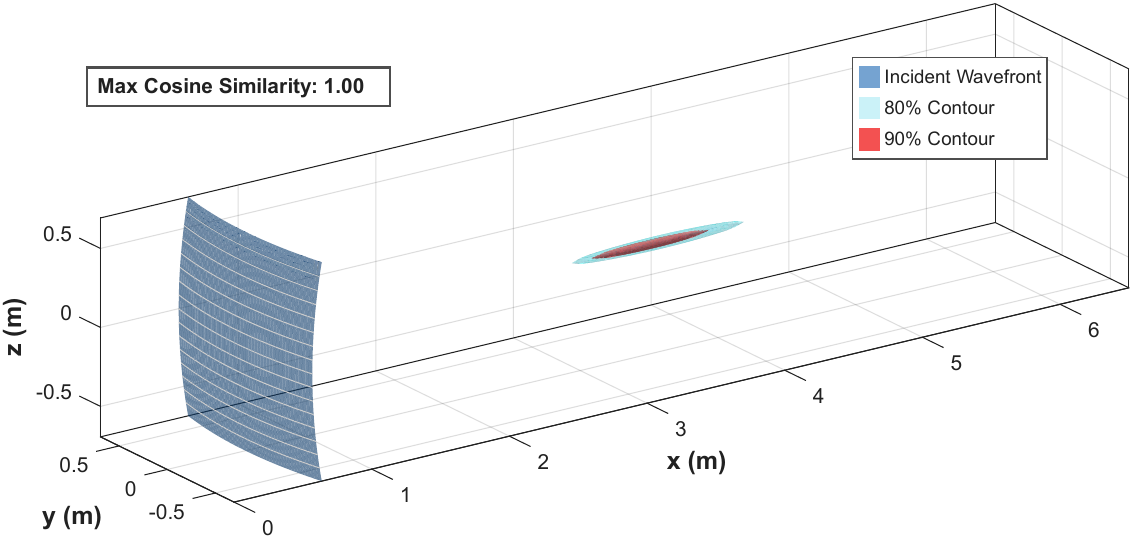} \\ 
    \includegraphics[width=\linewidth]{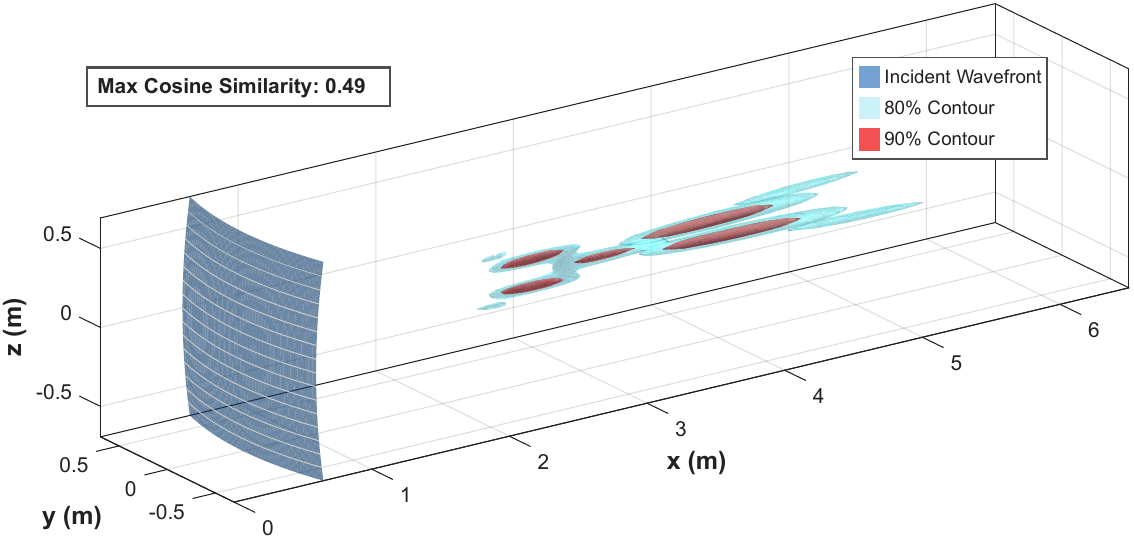}
    \caption{Spatial similarity distribution of (top) SWC and (bottom) AWC. For clarity, the BS array is omitted in the figure.}
    \label{fig:wavefront_comparison}
\end{figure}

In Fig. \ref{fig:wavefront_comparison}, we evaluate the degree to which the channels corresponding to two distinct incident electromagnetic waves can be fitted by polar-domain steering vectors. Both waves are incident normally on the array but exhibit different wavefront curvatures. For the AWC, the principal radii of curvature are 2.5 m and 4.5 m, with the principal directions aligned with the array axes. For the SWC, the radius of curvature is 3.5 m. The antenna array size is set to $64 \times 64$, the carrier frequency is 7.5 GHz, and the antenna spacing is a half-wavelength.

Fig. \ref{fig:wavefront_comparison} illustrates the non-sparsity of AWC in the polar domain by highlighting the maximum cosine similarity achieved by the polar-domain steering vectors for both channels, along with the volumetric contours capturing 80\% and 90\% of this peak similarity. As observed, the spatial similarity distribution of the SWC is highly concentrated, reaching a maximum of $1.0$ at $x = 3.5$ m along the propagation axis. This indicates that the SWC channel can be sparsely and accurately represented by a single point source. In contrast, a single point source can achieve a maximum cosine similarity of only about 0.5 with the AWC. Instead of a point focus, the volumetric energy of the AWC diffuses into two distinct, orthogonal caustic line segments located at $x = 2.5$ m and $x = 4.5$ m, corresponding to its two principal radii of curvature.

\subsection{Impact of Propagation Characteristics on AWC Anisotropy}

As detailed in Section \ref{sec: AWC Formulation}, the factors governing the anisotropic characteristics of the AWC encompass the distance from UE to the scatterer (which dictates the incident wavefront curvature at the reflection point), the distance from the scatterer to the BS array center, the BS array scale, the carrier wavelength and the surface curvature of the scatterer. In this subsection, we design the following experiments to investigate the impact of these propagation characteristics:

\begin{table*}[t]
  \centering
  \caption{NMSE of SWC-Fitting Across Various Physical Configurations}
  \label{tab:nmse_results}
  \resizebox{\textwidth}{!}{
  \begin{tabular}{|c|ccccc|c|ccccc|}
    \hline
    \multicolumn{6}{|c|}{\textbf{Scatterer-BS Distance}} & \multicolumn{6}{c|}{\textbf{UE-Scatterer Distance}} \\
    \hline
    $r$ (m) & 5 & 8 & 10 & 15 & 20 & $d$ (m) & 2.0 & 4.0 & 6.0 & 8.0 & 22.8 \\
    NMSE & 8.94e-01 & 8.06e-01 & 6.99e-01 & 3.48e-01 & 1.78e-01 & NMSE & 4.33e-02 & 1.42e-01 & 2.43e-01 & 3.30e-01 & 6.30e-01 \\
    \hline
    \multicolumn{6}{|c|}{\textbf{Array Scale \& Frequency}} & \multicolumn{6}{c|}{\textbf{Scatterer Curvature}} \\
    \hline
    $N$, $f_c$ & $32^2$, 3.8G & $64^2$, 7.5G & $128^2$, 15.0G & $192^2$, 22.5G & $256^2$, 30.0G & $R^{-1}$ (m$^{-1}$) & 0.0 & 0.5 & 2.0 & 4.0 & 10.0 \\
    NMSE & 6.22e-02 & 2.27e-01 & 6.30e-01 & 8.29e-01 & 8.88e-01 & NMSE & 1.06e-12 & 4.17e-01 & 5.95e-01 & 6.30e-01 & 6.50e-01 \\
    \hline
  \end{tabular}
  }
\end{table*}

\begin{itemize}
    \item Varying the distance from the scatterer to the BS array center, while fixing the BS-scatterer relative direction and the UE-scatterer relative position.
    \item Varying the distance from the UE to the scatterer, while maintaining the UE-scatterer relative direction and the BS-scatterer relative position constant.
    \item Varying the antenna array size and the carrier wavelength simultaneously, while keeping the array aperture constant.
    \item Varying the surface curvature of the scatterer.
\end{itemize}

Unless otherwise specified, the default parameters for the aforementioned experiments are set as follows: the array size is $128 \times 128$, and the carrier frequency is $15$ GHz. The center of the BS array is located at $[0, 0, 10]^T$ m, and the UE is located at $[30, 0, 1.5]^T$ m. The propagation scenario comprises a single cylindrical scatterer, whose principal curvatures are $4\text{ m}^{-1}$ and $0\text{ m}^{-1}$ aligned with the $x$- and $z$-axes, respectively. For each generated AWC in the experiments, we search for an optimal SWC that minimizes the normalized mean square error (NMSE) between the two channel vectors. This NMSE is then employed as the quantitative metric to characterize the degree of wavefront anisotropy of the AWC.

The simulation results in Table \ref{tab:nmse_results}, reveal how propagation characteristics influence the degree of wavefront anisotropy:

\begin{itemize}
\item \textbf{Scatterer-to-BS distance:} A larger distance reduces anisotropy. As the scattered wave travels, its wavefront naturally flattens. This reduces the difference between its two principal curvatures, making it more like an SWC.
\item \textbf{UE-to-scatterer distance:} A shorter distance also reduces anisotropy. A closer UE creates a larger incident curvature on the scatterer, giving the reflected wave a larger initial curvature. For an anisotropic wavefront, the direction with the larger curvature flattens faster. A larger initial curvature increases this difference in flattening speeds between the two directions, thereby reducing the wavefront anisotropy when it reaches the BS.
\item \textbf{Carrier frequency and array size:} For a fixed physical aperture, a higher frequency and a larger array scale makes the channel phase more sensitive to the wavefront shape. This makes the effect of anisotropy more obvious.
\item \textbf{Scatterer surface curvature:} A larger surface curvature directly makes the scattered wave more anisotropic from the start, causing a larger difference from the SWC model. Specifically, when the curvatures of the scatterer in both directions are zero, the reflection is specular. This is equivalent to an electromagnetic wave emitted from a virtual source of the UE directly reaching the BS, making 
AWC identical to an SWC.
\end{itemize}
\subsection{Performance of the Proposed Algorithm}

In this subsection, we evaluate and compare the NMSE performance of the following algorithms and theoretical bounds in the AWC and SWC scenarios, respectively:
\begin{itemize}
    \item \textbf{AWC-Estm:} The proposed AWC model-based channel estimation algorithm.
    \item \textbf{SWC-OMP-LM:} A baseline algorithm that utilizes a dictionary constructed from standard spherical wave steering vectors for initial channel estimation via OMP, followed by a LM algorithm for parameter refinement. The OMP path extraction terminates when the reduction in the residual falls below 5\% of the reduction achieved in the previous iteration.
    \item \textbf{CRLB:} The Cram\'er-Rao Lower Bounds derived based on the two respective channel models, denoted as CRLB(AWC) and CRLB(SWC).
\end{itemize}

\begin{figure}[htbp]
    \centering
    \includegraphics[width=\linewidth]{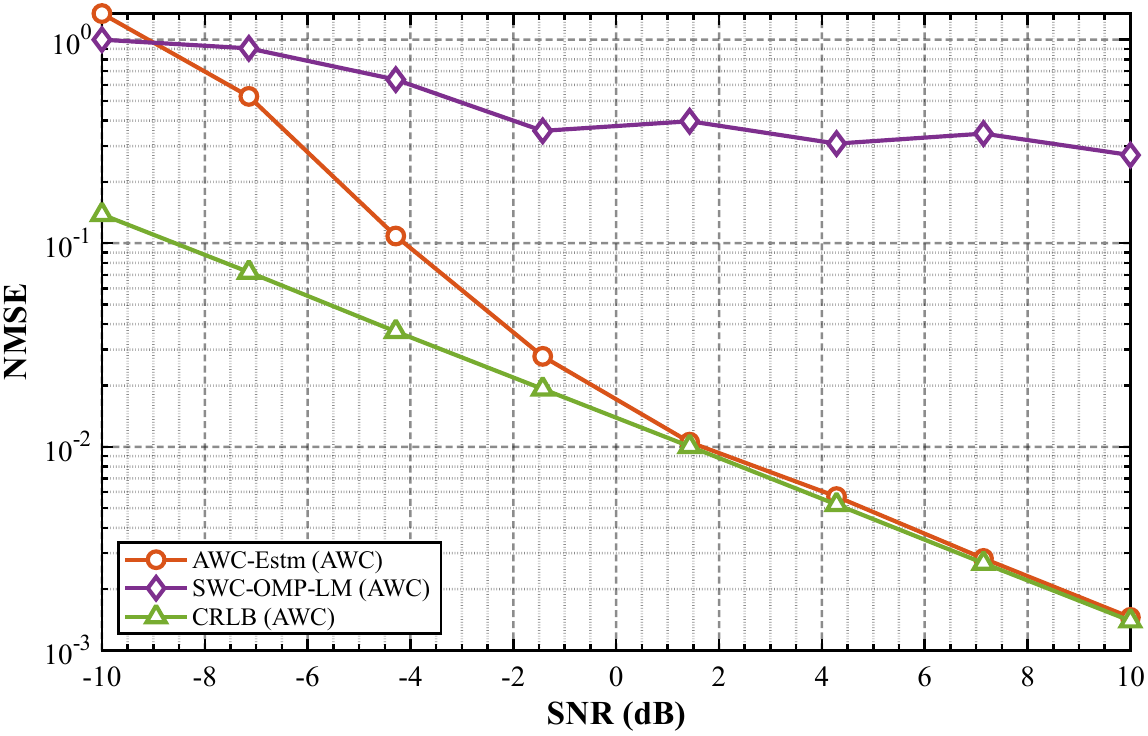} \\ 
    \includegraphics[width=\linewidth]{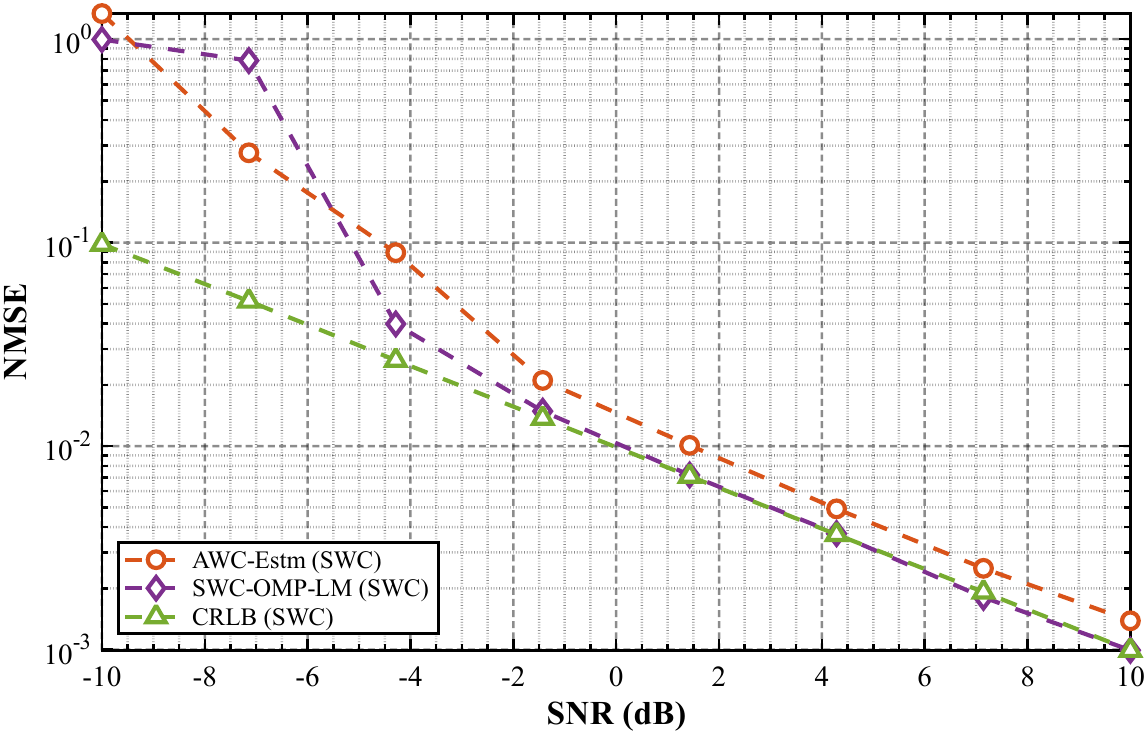}
    \caption{NMSE curves for (top) AWC and (bottom) SWC.}
    \label{fig:NMSE_vs_SNR}
\end{figure}

The propagation scenario for generating the AWC aligns with the default parameters established in the previous subsection, with the exception that the scatterer's location is randomly generated between the UE and the BS, and its principal curvature directions are also randomly oriented. The scenario for generating the SWC is identical to that of the AWC, except that both curvatures of the scatterer are set to zero. $N_\text{RF}$ and the pilot length $P$ are both set to $16$. The NMSE results are obtained by averaging over 100 independent random scenarios. Fig. \ref{fig:NMSE_vs_SNR} presents the comparative NMSE curves evaluated in AWC and SWC scenarios, respectively.

In the AWC environment, the proposed algorithm approaches the CRLB at high SNRs, which validates its effectiveness. At low SNRs, a performance gap exists relative to the CRLB because the phase differencing operation amplifies the noise, which yields false peaks on the spectrum of the differential matrices $\boldsymbol{\Psi}_1,\boldsymbol{\Psi}_2$ and degrades the subsequent parameter extraction. However, the proposed method still significantly outperforms the SWC-OMP-LM baseline across nearly all SNR levels. The accuracy of SWC-OMP-LM is limited by the modeling mismatch between its SWC dictionary and the actual AWC. This mismatch creates an error floor, meaning the NMSE stops improving even if the SNR continues to increase.

Conversely, in the SWC environment, the proposed algorithm shows a slight performance loss compared to the CRLB at high SNRs. This gap is caused by overparameterization. Since our method is designed based on the general AWC model, it must estimate a full curvature matrix rather than a single scalar curvature. The two redundant parameters in this matrix introduce additional estimation variance. In contrast, for SWC-OMP-LM there is no modeling mismatch in this scenario, thus it can accurately extract the channel parameters and its NMSE converge to the CRLB.

\section{Conclusion}\label{sec: Conclusion}
This paper investigates near-field channels exhibiting wavefront anisotropy in ELAA-mmWave systems. We first establish a parameterized model for such channels and subsequently design a channel estimation algorithm for the AWC. Through simulations, we investigate the non-sparsity of the AWC in the polar domain, reveal the impact of propagation scenario characteristics on the degree of wavefront anisotropy, and validate the effectiveness of the proposed algorithm.

Simulation results demonstrate that the energy of AWC in the polar domain disperses into specific regions determined by its principal curvatures and their corresponding principal directions, rendering it incapable of being represented by a single spherical wave. Furthermore, a shorter scatterer-to-BS distance, a larger UE-to-scatterer distance, a shorter wavelength, and a larger scatterer surface curvature all exacerbate the wavefront anisotropy. In such anisotropic environments, conventional SWC-based estimation algorithms suffer from severe performance degradation due to modeling mismatch, whereas our proposed algorithm successfully approaches the CRLB. Meanwhile, in traditional SWC scenario, the overparameterization of our algorithm leads to only a slight performance degradation relative to the CRLB. Future research will extend this work to the modeling and estimation of wideband multipath AWCs.

\end{document}